# Prospectively accelerated dynamic speech MRI at 3 Tesla using a self-navigated spiral based manifold regularized scheme

*Rushdi Zahid Rusho[1], Abdul Haseeb Ahmed[2], Stanley Kruger[3], Wahidul Alam[1], David Meyer[4], David Howard[5], Brad Story[6], Mathews Jacob[2], Sajan Goud Lingala[1,3]*

[1] *Roy J Carver Department of Biomedical Engineering, University of Iowa, Iowa city, IA, 52242*

[2] *Department of Electrical and Computer Engineering, University of Iowa, Iowa city, IA, 52242*

[3] *Department of Radiology, University of Iowa, Iowa city, IA, 52242*

[4] *Janette Ogg Voice Research Center, Shenandoah University, Winchester, VA, 22601*

[5] *Department of Electronic Engineering, Royal Holloway, University of London, UK*

[6] *Department of Speech, Language, and Hearing Sciences, University of Arizona, AZ, 85721*

Corresponding author
Sajan Goud Lingala, PhD
Assistant Professor
Roy J Carver Department of Biomedical Engineering
Department of Radiology
University of Iowa
sajangoud-lingala@uiowa.edu

Approximate word count: 5652
Figures: 10
Supplementary material: 4 figures, and 14 videos.

**ABSTRACT:** This work develops and evaluates a self-navigated variable density spiral (VDS) based manifold regularization scheme to prospectively improve dynamic speech MRI at 3 Tesla. Short readout duration spirals (1.3 ms long) were used to minimize sensitivity to off-resonance. A custom 16-channel speech coil was used for improved parallel imaging of vocal tract structures. The manifold model leveraged similarities between frames sharing similar vocal tract postures without explicit motion binning. The self-navigating capability of VDS was leveraged to learn the Laplacian structure of the manifold. Reconstruction was posed as a SENSE-based non-local soft weighted temporal regularization scheme. Our approach was compared against view-sharing, low-rank, finite difference, extra-dimension-based sparsity reconstruction constraints. Under-sampling experiments were conducted on five volunteers performing repetitive and arbitrary speaking tasks at different speaking rates. Quantitative evaluation in terms of mean square error over moving edges were performed in a retrospective under-sampling experiment on one volunteer. For prospective under-sampling, blinded image quality evaluation in the categories of alias artifacts, spatial blurring, and temporal blurring were performed by three experts in voice research. Region of interest (ROI) analysis at articulator boundaries were performed in both the experiments to assess articulatory motion. Improved performance with manifold reconstruction constraints was observed over existing constraints. With prospective under-sampling, a spatial resolution of 2.4mm$^2$/pixel and a temporal resolution of 17.4 ms/frame for single slice imaging, and 52.2 ms/frame for concurrent 3-slice imaging were achieved. We demonstrated implicit motion binning by analyzing the mechanics of the Laplacian matrix. Manifold regularization demonstrated superior image quality scores in reducing spatial and temporal blurring compared to all other reconstruction constraints. While it exhibited faint (non-significant) alias artifacts similar to temporal finite-difference, it provided statistically significant improvements over remaining constraints. In conclusion, the self-navigated manifold regularized scheme enabled robust high spatio-temporal resolution dynamic speech MRI at 3 Tesla.

Abbreviations:
MRI: Magnetic Resonance Imaging
VDS : variable density spiral
SENSE: sensitivity encoding
L matrix: the graph Laplacian manifold matrix
ROI: Region of interest
TFD: Temporal Finite difference
XD-sort: Extra dimension sorting based sparsity regularization
FOV: Field of View
GRE: Gradient echo
TR: Repetition time
TE: Echo time
SNR: Signal to noise ratio
BM4D: Block matching 4D
MSE: mean square error

**INTRODUCTION**

Speech production involves an intricate coordination of several soft tissue structures including the lips, tongue, soft-palate, epiglottis, vocal folds, and the pharyngeal wall. Safely visualizing the dynamics of both the deforming vocal tract air space and the neighboring articulators is a powerful means for exploring and understanding better the subtle nuances of speech production. Dynamic magnetic resonance imaging has emerged as the modality of choice to visualize speech because of several advantages over competing modalities such as lack of ionizing radiation, ability to capture longitudinal measures, flexibility to image in arbitrary image planes, and the ability to visualize deep structures such as the vocal folds [1,2]. It has been used in several speech science and clinical studies to understand better apraxia of speech [3], quantifying levator veli palatini muscle movements [4], answering open questions in phonetics (eg. understanding timing effects of nasal syllables [5]), and to understand better speech production in different languages such as French [6,7], Arabic [8], Brazilian Portugese [9]. It has also been applied in studies aimed to understand mechanics of vocal tract shaping during singing [10,11], beatboxing [12,13], playing musical instruments such as the horn [14].

In the past decade, several schemes based on sparse k-space v.s time (k-t) sampling and model-based reconstruction have been applied to improve spatio-temporal resolutions, and/or vocal tract coverage at both 1.5 T and 3 T [9,14–20]. Several constrained models have been applied, including those that exploit transform sparsity [14–17], joint low rank and transform sparsity [20,21], and low rank plus transform sparsity [19]. However, these constraints can be susceptible to motion blurring and loss of spatial and temporal features while modeling fast arbitrary articulatory motion during speech. Spiral trajectories with short readouts ($\leq$ 2.5 ms) have been the natural choice at 1.5 T due to their robustness to encode fast motion [1]. However, at 3 T, adapting a 2.5 ms duration readout can induce more than 2 cycles of phase accrual, and cause significant blurring at air-tissue boundaries. It is noteworthy that, off-resonance can be significant at 3T, up to ~1200 Hz. Lesser time efficient radial trajectories are more commonly used due to their short readouts (~1.2 to 1.4 ms) [1]. Under-sampled Cartesian trajectories, combined with 3D spiral navigators, and a reconstruction employing joint low rank sparsity constraints have been used for improved 3D dynamic rapid imaging at [20–22]. However, these schemes are challenged by long acquisition times (eg. of the order of 7.5 to 20 minutes), which may introduce subject fatigue during natural speech production.

Explicit motion binning strategies have shown significant promise to reduce motion blurring associated with classic low rank, and/or sparsity models [23–25]. These models have largely been applied in free breathing applications (eg. free breathing abdomen MRI). These models rely on navigator data such as those from respiratory bellows, or data derived from k-t space to extract the breathing motion pattern. These data are then used to organize the dynamic time series into image frames with similar motion state. The models benefit by imposing sparsity/low rank assumption within each motion state, i.e, along this "extra dimension." However, performing such binning is infeasible in dynamic speech MRI because the motion can be arbitrary. An explicit deformation estimation and correction scheme was applied to improve the low rank and sparsity model in 3D dynamic speech MRI [26]. However, joint deformation estimation and reconstruction schemes face practical challenges including sensitivity to local minima solutions and the introduction of non-trivial interpolation artifacts while correcting for large deformations [27].

Emerging manifold regularization approaches have shown significant promise to improve motion fidelity in free breathing and ungated cardiac MRI [28–30]. These schemes model image frames as points living on a low dimensional manifold embedded in a high dimension space. Similar image frames are mapped as neighbors on the manifold, even if they are distant in time. A key difference with extra dimension schemes is that it *implicitly exploits similarities between image frames without the need of explicit motion binning or explicit deformation estimation*. This makes it an attractive method to explore in the context of arbitrary speech production where explicit binning is not feasible. In the original manifold regularized MRI framework [31], the authors demonstrate a dynamic speech MRI example, showing improved reconstruction quality with the manifold constraint against a low rank only, and a sparsity only constraint. However, this single example is very restrictive, as it was based on retrospective down-sampling of view-shared reconstructed dynamic speech data acquired at 1.5 T. To date, there is no study demonstrating the prospective utility of manifold reconstruction constraints for improving dynamic speech MRI at 3 T.

In this work, we aim to accelerate prospectively dynamic speech at 3 T by using implicit motion resolved capabilities of manifold regularization. We also synergistically combine parallel imaging capabilities of a recently developed custom 16-channel vocal tract receive coil [32], and self-navigation capabilities of short readout variable density spirals. Variable density spirals that oversample the center of the k-space are used to estimate a graph Laplacian matrix, which determines the neighborhood relations between image frames. To minimize off-resonance artifacts, our spirals are implemented with extremely short readouts (~ 1.3 ms), which is similar to readout duration of widely used radials at 3 T. The reconstruction is posed as a SENSE-based $l_2$ norm non-local soft weighted temporal regularization scheme. We demonstrate the applicability of our approach to enable single slice (mid sagittal) imaging at 17.4 ms/frame; and concurrent three slice (1 mid, and 2 parasagittal) imaging at 52.2 ms/frame at a spatial resolution of 2.4 mm²/pixel. A variety of speech tasks were imaged, including repeatedly producing vowel and consonant sounds, and producing fluent speech by five healthy volunteer speakers. We analyze the mechanics of the Laplacian matrix in these speech tasks, and qualitatively demonstrate implicit binning enabled by our approach. We compare our proposed reconstruction against view-sharing, low-rank, finite difference, and extra dimension-based sparsity reconstruction constraints. We performed both retrospective and prospective under-sampling experiments. Quantitative evaluation of mean square error over moving edges were performed in the retrospective under-sampling experiment. For prospective under-sampling, blinded image quality evaluation of alias artifacts, spatial blurring, and temporal blurring were performed by three experts in voice research. In both retrospective and prospective under-sampling experiments, region of interest (ROI) analysis at articulator boundaries were performed to assess articulatory motion over a speech token.

**METHODS**

Figure 1 shows the manifold model schematic, where dynamic images are modeled as points on a low dimensional manifold embedded in a high dimensional ambient space. The dimension of the ambient space equals the number of pixels in the image. Image frames with a similar anatomical vocal tract posture are mapped as neighbors on the manifold, even if they occur arbitrarily in time. Manifold regularization exploits similarity of image frames in terms of proximity of points on the smooth manifold. The recovery is posed as an optimization problem with the criterion [33]:

$$\|\mathbf{A}(\mathbf{X}) - \mathbf{b}\|_2^2 + \lambda\, trace(\mathbf{X}\mathbf{L}\mathbf{X}^{\mathbf{H}}); \quad (1)$$

In this equation, **A** is an operator modeling coil sensitivity encoding and Fourier Transform operation on a spiral trajectory. The Casorati matrix containing the dynamic image time series (whose columns are the image frames) is represented as $\mathbf{X}_{M \times N} = [x_1; x_2; ..; x_N]$. *N* represents the number of frames, and *M* represents total number of pixels in each frame. **b** is a vector containing the acquired under-sampled multi-coil k-space data for all time frames. In this work, we use a recently proposed custom 16-channel vocal tract receiver coil [32]. The custom coil provides high spatial sensitivity over various regions of interest including tongue, soft palate, pharyngeal wall, larynx, and infra glottic airspace. The graph Laplacian matrix ($\mathbf{L}_{N \times N}$) captures the structure of the manifold, and is defined as $\mathbf{L}_{N \times N} = \mathbf{D} - \mathbf{W}$; where **W** is the $N \times N$ weight matrix containing the weights $w_{ij}$; **D** is the diagonal matrix, with diagonal entries, $D_{ii} = \sum_j w_{ij}$. $\lambda$ is a tunable parameter that balances data consistency and regularization. The above optimization criterion is easier to understand intuitively and can also be expressed as:

$$\|A(\mathbf{X}) - \mathbf{b}\|_2^2 + \lambda \sum_{i=1}^{N}\sum_{j=1}^{N} w_{ij}\|x_i - x_j\|_2^2; \quad (2)$$

Note that the second term comprises of a weighted penalty where the weights $w_{ij}$ determine the degree of similarity between the $i^{th}$ and the $j^{th}$ frame; and are inversely proportional to the distance between the corresponding points on the manifold. That is to say, similar frames are assigned higher weights, and vice-versa.

The second term above now may be viewed as a $l_2$ norm based non-local soft-weighted temporal regularization operation. Each row of the L matrix would reveal a $1 \times N$ kernel containing coefficients used for that corresponding frame. Note, that the L matrix is derived from the data itself (discussed below). This is in stark contrast to the widely used local temporal finite difference kernel, where the L matrix would be a predetermined block diagonal matrix with entries [-1,2,1].

*Variable density spiral design:*

We designed a variable density spiral (VDS) trajectory that allows for self-navigation [34]. The design parameters were: slew rate = 140 mT/m/ms; maximum gradient amplitude = 80 mT/m; sampling time = 4 µs; spatial resolution = 2.4 mm x 2.4 mm; 27 spiral arms. When the normalized k-space radius ($k_r$) was between (0 to 0.25); (0.25 to 0.5); (0.5 to 1), the FOV respectively changed linearly from (60 cm² to 30 cm²); (30 cm² to 20 cm²); and (20 cm² to 6.66 cm²). With this design, VDS produced 1.3 ms long readouts with 335 readout points. For the single slice setting, successive spiral arms were interleaved by the golden angle (~222.379°). For concurrent multi-slice setting, golden angle increments occurred only after spiral arms at a specified angle were acquired from all the slices [15].

*Estimating the Laplacian matrix:*

In this study, we exploit self-navigating capability of the VDS trajectory to estimate the **L** matrix. We denote the central $q\%$ of the VDS k-space samples as $\mathbf{b}_{low-res}$. Starting with an initial guess of $x_{low-res}$ obtained from the inverse nuFFT of $\mathbf{b}_{low-res}$, we determine **L** by iterating through the following steps:

$$w_{ij} = \exp\left(-\|x_{low-res,i} - x_{low-res,j}\|_2^2 / \delta^2\right); \quad (3)$$

$$\mathbf{L}_{NxN} = \mathbf{D} - \mathbf{W}; \quad (4)$$

$$min_{\mathbf{X}_{low-res}} \|A(\mathbf{X}_{low-res}) - \mathbf{b}_{low-res}\|_2^2 + \lambda\, trace(\mathbf{X}_{low-res} \mathbf{L} \mathbf{X}_{low-res}^H);\quad (5)$$

The weights in (3) are determined through the Gaussian Kernel, where $\delta^2$ is a tunable parameter that controls the smoothness of the manifold. $x_{low-res}$ is updated by solving (5) using a conjugate gradient algorithm. We iterate between (3) to (5) to ensure the **L** matrix is learned from artifact free data, as $\mathbf{b}_{low-res}$ typically contain few missing samples in the highly under-sampling scenario. The number of iterations were set to 5.

*Reconstruction:*

Once the **L** matrix is obtained, one can perform manifold regularized reconstruction of dynamic images by solving (1). However, for faster processing, we performed an eigen decomposition of **L** as $\mathbf{L} = \mathbf{V}\Sigma\mathbf{V}^H$, and used the $r$ smallest eigen vectors $\mathbf{V}_{Nxr}$ to approximate $\mathbf{X} = \mathbf{U}\mathbf{V}^H$. The matrix of r spatial basis images $\mathbf{U}_{Mxr}$ were estimated by solving the following computationally simpler optimization problem using a nonlinear conjugate gradient algorithm [35]:

$$\mathbf{U}^* = \min_{\mathbf{U}} \|\mathbf{A}(\mathbf{U}\mathbf{V}^H) - \mathbf{b}\|_F^2 + \lambda \sum_{i=1}^{r} \sigma_i \|\mathbf{u}_i\|^2 \;;\quad (6)$$

The final Casorati matrix was then obtained as $\mathbf{X} = \mathbf{U}\mathbf{V}^H$.

Reconstruction was implemented in MATLAB (The MathWorks, Inc., Natick MA) on a high-performance computing cluster at The University of Iowa, equipped with an Intel Xenon central processing unit with 28 cores at 2.40 GHz and 128 GB of memory, and a NVIDIA Tesla P100-PCIE graphical processing unit with 16GB memory. Three coil-elements that captured artifact energy from inferior heart anatomy outside the vocal tract FOV of interest, and were omitted in the reconstruction. Raw k-space data from the remaining 13 coils were coil compressed to 8 virtual coils via PCA-based coil compression. The gpu NUFFT function was used to implement nuFFT operations in **A** [36]. Virtual coil sensitivity maps were estimated from time averaged data using the E-SPIRIT coil map estimation algorithm [37].

*In-vivo experiments:*

Five healthy adult volunteers (4 male; 1 female; ages 27-50) were scanned on a GE 3 T Premier scanner equipped with high performance gradients. The study was approved by University of Iowa's institutional review board (IRB: 200810706), where a full explanation was given with opportunities for interaction and written consent was obtained from all subjects before scanning. In order to obtain a fully-sampled dataset for our retrospective under-sampling experiments, we scanned one of the volunteers with a Cartesian GRE acquisition in the mid-sagittal orientation with the following parameters: flip angle = 5 degrees; slice thickness = 6mm; number of phase and frequency encodes = 68 x 68; field of view = 200 mm x 200 mm; TR = 2.58 ms; TE = 1.17 ms; receiver bandwidth = $\pm$ 125 kHz; temporal resolution = 183 ms; acquisition time = 94 seconds. The subject performed speaking tasks of producing interleaved consonant and vowel sounds, and counting numbers indefinitely. In this acquisition, the subject voluntarily produced speech at a substantially slower speaking rate than his natural pace. This acquisition also compromised the spatial resolution to 2.94 mm$^2$. Together, this ensured capturing major articulatory movements at the coarse 183 ms temporal resolution. Coil sensitivity maps were estimated from time averaged data using the root sum of squares approach. Coil combined image space data was then generated by using the R=1 SENSE multi-coil combination. The dynamic images had a low signal to noise ratio (SNR) due to the use of a high bandwidth acquisition. To improve the SNR, we applied a block matching 4D (BM4D) spatio-temporal denoising algorithm [38], and have generated the ground truth reference image space dataset. To simulate retrospective under-sampling, for every image frame on the reference image space data, we applied the forward multi-coil nuFFT model operator with VDS sampling trajectory to generate the under-sampled multi-coil k-space vs time data. We performed experimentation for under-sampling factors corresponding to 4 spiral arms/frame, 3 spiral arms/frame, and 2 spiral arms/frame.

For all five volunteers, prospective under-sampling experimentation was performed with the variable density spiral based gradient echo sequence with rewinding. Imaging parameters were TR=5.8 ms, flip angle=5$^0$; total number of spiral interleaves = 2700; slice thickness = 6mm; spatial resolution = 2.4 mm$^2$. Single slice acquisitions were performed in the mid-sagittal plane. Concurrent three slice acquisitions were performed in 1 mid-sagittal, and 2 para-sagittal planes. Reconstruction was performed by combining every 3 interleaves/frame, which corresponded to a time resolution of 17.4 ms/frame for the single slice setting; and 52.2 ms/frame for the concurrent three slice setting. The volunteers produced the following speaking tasks: a) repetition of the phrase "za-na-za"; b) repetition of the phrase "loo-lee-la-za-na-za" ; c) producing arbitrary/fluent speech by counting numbers (starting from "one") indefinitely through the scan duration. Note tasks in a) and b) involved interleaving of consonant and vowel sounds. The above speech tokens in (a) and (b) are respectively represented as /za na za/ and /lu li la za na za/ under the international phonetic alphabet (IPA) notation.

In both the retrospective and prospective under-sampling experiments, the k-space vs time data were reconstructed with five different reconstruction algorithms: 1) a view-sharing based scheme which is commonly used in dynamic speech MRI [39], 2) a nuclear norm-based low rank regularization scheme [40],

3) an $l_1$ norm sparsity-based temporal finite difference regularized scheme (TFD) [15], 4) an extra dimension sorting based sparsity regularization scheme (XD-sort) [23], and 5) the proposed manifold regularization scheme. The view-sharing reconstruction scheme formed images for every frame by combining 27 arms/frame, and used a view-step size corresponding to the target temporal resolution (eg. 3 spiral interleaves in the prospective setting). The extra dimension based sparsity regularization scheme was implemented using a sorting strategy proposed in XD-GRASP [23]. Briefly, a singular value decomposition was performed on the converged low-resolution $\mathbf{X}_{low-res}$ images from eq. (5). The first significant temporal singular vector was extracted and sorted in an ascending order. The sorted indices were then used to sort the frames of the under-sampled k-space v.s time dataset. This frame-sorted k-space v.s time under-sampled data was then reconstructed using the temporal finite difference reconstruction, after which the final reconstruction was reverse sorted to represent their actual time of occurrence. The sparsity-based finite difference regularized scheme, the low-rank scheme, and the extra dimension-based regularization scheme were implemented in the Berkeley advanced reconstruction tool box (BART) computing environment on the above Intel Xenon CPU [41]. The dynamic Casorati matrix ($\mathbf{X}$) to be reconstructed was of the size $168^2$x900. Multiple slices in the concurrent three slice acquisition were reconstructed independently. Reconstruction times for the manifold regularized scheme was of the order of 94 minutes, with the L matrix estimation step being implemented on the CPU which took ~83 minutes; and the U matrix estimation step was implemented on the GPU, which took ~11 minutes., Reconstruction times for the temporal finite difference, low rank, and extra dimension-based sparsity regularization were of the order of 14.5-18 minutes

*Tuning free parameters in the reconstruction:*

The proposed manifold regularized reconstruction depends on four parameters: i) the central q% k space samples in $\mathbf{b}_{low-res}$; ii) manifold smoothness parameter ($\delta^2$) in (3); iii) regularization parameter ($\lambda$) in (6); iv) number of eigen basis functions (*r*) during the eigen decomposition of the $\mathbf{L}$ matrix. In the retrospective under-sampling experiments, we determined these parameters by optimizing the mean square error between the reconstructions and the reference datasets over the vocal tract regions of interest. In the prospective under-sampling experiments, we empirically determined these parameters by visually assessing tradeoffs amongst aliasing artifacts, spatial and temporal blurring. We note this parameter choice was in accordance with the L-curve heuristic that determines the best regularization parameter balancing energy between data consistency and regularization. We performed this experiment on two datasets from two representative speakers (fluent speech and repetitive speech tasks), and determined the parameters as: q=18%; $\delta^2 = 4.5$; $\lambda = 0.2$; $r = 30$. Image quality for these representative datasets for different parameters is included in the supplementary material. We observed this choice to be robust across all the datasets from the five speakers. We used the same criterion as above to determine the regularization parameters in the finite difference regularization, low rank regularization, and extra-dimensional regularization schemes.

*Region of interest (ROI) time profile analysis:*

To examine vowel and consonant articulation during speech task, "loo-lee-la-za-na-za", a region of interest (ROI) analysis to analyze the constrictions of the vocal tract air space in both the retrospective and prospective under-sampling experiments [42]. We considered three different ROIs along the vocal tract in the midsagittal reconstruction of the 3-slice acquisition from one representative subject. These were at the landmarks of a) apex (tip) of the tongue; b) lower lip; and c) velum. A mask was drawn to enclose each of these ROIs and kept fixed for all the reconstructed image frames in the dataset. We then calculated the mean of the pixel intensities within the ROI for each image time frame, and visualized it as a one-dimensional time varying signal. These ROI time profiles were compared amongst all the reconstructions.

*Visualizing the mechanics of the Laplacian matrix:*

To demonstrate the implicit binning enabled by manifold regularization, we visualized the Laplacian matrix for two different speech tasks from one of the volunteers in the prospective under-sampling dataset: a) fluent task of counting numbers indefinitely; and b) repetitive speaking task of repeating the phrase "loo-lee-la-za-na-za". Here, we considered representative reconstructions from 3 slice acquisition for the fluent task, and single slice acquisition for the repetitive speaking task. We analyzed representative rows of the Laplacian matrix, which revealed underlying non-local kernel coefficients used during the proposed non-local weighted temporal regularization. Similar vocal tract postures from different frames corresponding to the peaks of the weights were analyzed. For a representative single slice reconstruction at 17.4 ms/frame, the temporal and spatial bases matrices, V and U were visualized and contrasted for the fluent counting and repetitive speaking tasks. These bases were plotted for the first five coefficients.

*Image quality assessment by expert raters:*

The prospective under-sampled reconstructions from the five reconstruction algorithms (view-sharing, low-rank regularization, temporal finite difference regularization, extra-dimension sparsity based regularization, and manifold regularization) were evaluated in the categories of aliasing artifacts, spatial blurring, and temporal blurring. Three experts in voice research (co-authors: D. Meyer, D. Howard, and B. Story) performed blinded ratings of the images. The scoring criterion in different categories followed a four-point scale:

Alias artifacts category:

1 - unacceptable; strong alias artifacts present and hampers visualization of all articulators.

2 - adequate; moderate level alias artifacts, and moderate level of interference with articulator visualization.

3 - good; faint alias artifacts present but does not hamper interpretation of articulatory motion

4- excellent; no alias artifacts.

Spatial blurring category:

1 - unacceptable; strong blurring at air-tissue interfaces.

2 - adequate; moderate level blurring; the air-tissue boundaries are blurred and hampers interpretation of articulator boundaries.

3 - good; faint blurring present but does not hamper interpretation of articulator boundaries.

4- excellent; no perceivable blurring.

Temporal blurring category:

1 - unacceptable; strong motion artifacts/blurring; articulatory motion completely blurred.

2 - adequate; moderate level motion artifacts/blurring especially when articulator position change.

3 - good; faint motion artifacts present but does not hamper interpretation of articulatory motion

4- excellent; no perceivable motion blurring/artifacts.

A total of 14 datasets from each reconstruction algorithm were scored by each expert. In each dataset, the reconstructions from each algorithm were randomly shuffled and presented simultaneously to the expert rater. Datasets 1 to 5 consisted of reconstructions from the 3-slice acquisition, respectively for the five speakers who repeated the phrase "loo-lee-la-za-na-za". Similarly, datasets 6 to 10 consisted of reconstructions from 3-slice acquisition, respectively for the five speakers who produced the fluent counting task without repetition. Datasets 11 to 12 consisted of reconstructions from single slice acquisitions from the first two speakers producing the phrase "za-na-za". Finally, datasets 13 to 14 contained single slice reconstructions from the first two speakers producing the fluent counting task without repetition. One of the volunteers had a dental implant, which resulted in signal nulling near the tongue tip (Datasets 5 and 10) in all the reconstruction methods. The raters were therefore asked to perform the rating by ignoring this signal null artifact, and were asked to provide ratings by focusing on remaining parts of the image.

Non-parameter pairwise statistical comparison of the image quality scores were performed using the Kruskal-Wallis test. For the scores in the alias-artifact, temporal blurring, and spatial blurring categories, pair-wise comparisons were established between the manifold regularized reconstruction and the remaining four other reconstruction algorithms.

**RESULTS**

Figure 2 shows reconstructions from all the algorithms in the retrospective under-sampling experiment. The reference fully-sampled dataset is shown in the top row. Reconstructions corresponding to different under-sampling rates when using 2,3,4 spiral arms/frame are shown. The proposed manifold regularized approach provides improved spatial and temporal fidelity compared to other algorithms. Specifically, the low-rank regularized reconstruction depicts greater alias artifacts with increasing under-sampling factor. View-sharing and XD-sort algorithms demonstrate significant temporal blurring. The temporal finite difference regularized reconstruction depicts motion blurring in terms of temporal stair-casing artifacts. The improved reconstruction image quality in the manifold regularized scheme is also highlighted by the improved mean square error (MSE) values in the vocal tract regions of interest.

Figure 3 shows the region of interest time profile analysis on the 3 arm/frame reconstructions in the retrospective under-sampling experiment. Articulatory temporal dynamics corresponding to the speech utterance "za-na-za", were analyzed. Averaged ROI time profiles at the boundaries of tongue tip-airway, lower lip-airway, and velum-airway were analyzed. While all existing reconstruction algorithms show deviations of these time profiles with respect to the ground truth time profiles, we observed close correspondence of the time profile from the proposed manifold regularized reconstruction to the ground truth dataset.

Figures 4 and 5 shows the Laplacian matrix respectively for the speaking tasks of: repetition of the phrase of "loo-lee-la-za-na-za", and fluent task of counting numbers. Representative rows of the L matrix are shown which corresponded to two postures of the vocal tract anatomy, where the tongue was raised, and the tongue was lowered. For better visualization, we highlighted entries which were above 10% of the maximum entry in each row by the red color, and superimposed onto the remaining entries in blue color. For the repetitive task in figure 4, the rows of the Laplacian matrix revealed large entries that occurred in a quasi-periodic manner. For the fluent task in figure 5, the rows of the L matrix showed large entries that occurred less frequently, and arbitrarily in time. Note, the peaks in each row of the L matrix pointed to frames sharing similar vocal tract anatomy postures, even though they occurred distantly along the time dimension. This may be viewed as a soft weighted non-local temporal regularization operation.

Figure 6 shows the temporal bases (V) or the eigen vectors of the L matrix, and the estimated spatial coefficients (U) for the two speech tasks. We noted the temporal bases to be adaptive in nature, and captured the underlying motion dynamics of the speaking task. For the repetitive task of producing the phrase "za-na-za", we observed clear quasi-periodic dynamics in the bases. While for the fluent counting task, we observed more arbitrary motion patterns corresponding to free speech.

Figure 7 shows a representative spatial frame and image temporal profile from all the reconstruction schemes for the fluent speech task of counting numbers indefinitely acquired using the concurrent 3-slice acquisition scheme. Reconstructions were performed using 3 spiral arms/frame, which corresponded to ~52.2 ms/frame. We qualitatively observed the low rank-based reconstructions to depict unresolved aliasing artifacts, and spatio-temporal blurring that hampered the visualization of articulatory boundary movements. Reconstruction with temporal finite difference regularization removed aliasing in general but demonstrated temporal stair casing artifacts. This manifested as motion blurring, especially between frames where articulators changed their position (eg. raising of the tongue tip, opening of lips). View sharing and XD-sort reconstructions demonstrated considerable spatial and temporal blurring. The manifold reconstruction showed very subtle aliasing artifacts. However, it demonstrated minimal spatial or motion blurring compared to other schemes, and provided acceptable fidelity, capturing the arbitrary motion dynamics of the articulators.

Figure 8 shows qualitative comparisons with representative image frames of all the reconstruction algorithms from the 3-slice concurrent acquisition for the speech task of repeating the phrase "loo-lee-la-za-na-za". We show few example frames during the transition in and out of the /a/, /n/, /l/,/i/ sounds. Note, in comparison to the competing schemes, the proposed manifold regularization scheme provides crisper air-tissue boundaries at several vocal tract landmarks such as tongue tip, velar boundary, lower lip, tongue base. For example, the event of raising the tongue tip and hitting against alveolar ridge during the sound /n/ was significantly blurred or artifacted with the competing schemes, but robustly captured with the manifold regularization scheme.

Figure 9 shows three ROI-averaged time profiles from the mid sagittal reconstructed image of the 3-slice acquisition during the production of the sound /a/-/n/-/a/-/z/-/a/-/l/ for all the reconstruction algorithms. We observed poor depiction of articulatory motion in the low rank and XD-sort schemes. The temporal finite difference and view-sharing scheme showed reduced artifact level, but showed significant blurring at tongue tip, lips, and the velum boundaries. The manifold reconstruction produced clean depiction of the underlying motion patterns. For example, the tongue apex (tip) in ROI 1 and the lower lips in ROI 2 should be in the raised position near the hard palate at the beginning of the sounds /n/, /z/ and /l/ which is marked by a sharp increase of ROI time profile. This behavior is represented well in the manifold reconstruction but not in the other schemes. Similarly, the area between velum and airway in the ROI 3 changes very little for the sound /a/ which is depicted as flat lines in the ROI- averaged time profiles. However, unlike the pronunciation of sound /z/, pronunciation of sound /n/ involves the velum moving upward towards the nasal cavity, hence there is slight decrease in the area in ROI 3. During sound /z/, the tongue near the velum moves upwards, causing increased area in ROI 3. The manifold scheme successfully captured these articulatory patterns, but the other methods were not similarly successful.

Finally, reconstructions from all the algorithms for all the datasets in a video format are provided in the supplementary material.

Figure 10 shows the combined image quality expert ratings. Scores are displayed as a violin plot for each of the reconstruction schemes and for each categories. The median is highlighted by a circle and the interquartile range is shown as black vertical rectangle. The density of the violin plot at a particular score is proportional to the number of times that score was assigned. We also specify this number in the plots. We note consistent higher score distribution of the manifold scheme in the 3's (i.e, good) and 4's (excellent) in all the categories in comparison to other schemes. Using the Kruskal-Wallis test, we observed statistically significant differences in image quality scores of the manifold reconstruction compared to low-rank reconstructions, view-sharing, XD-sort, TFD reconstructions in the spatial blurring and temporal blurring categories ($p < 0.001$). In the alias artifact category, the statistically significant differences were maintained when the manifold reconstruction was compared to the low-rank, view-sharing, and XD-sort schemes ($p < 0.001$). TFD and manifold reconstructions showed minimal differences in image quality scores in the alias artifact category.

**DISCUSSION**

We demonstrated the feasibility of rapid dynamic speech imaging at 3 Tesla by synergistically combining the parallel imaging capabilities of a custom vocal tract coil, the self-navigation capabilities of short readout variable density spirals, and the implicit motion binning of manifold regularization. Our scheme prospectively achieved a temporal resolution of ~17.4 ms/frame for single slice and ~52.2ms/frame for concurrent three slice imaging at a spatial resolution of 2.4 mm$^2$/pixel. Our 27 arm VDS scheme employed a ~1.3 ms readout, which is similar to the readout duration of radials at 3 T. Through retrospective and prospective under-sampling-based experiments, we demonstrated improved reconstruction quality with our approach compared to competing algorithms such as view-sharing, low-rank, and temporal finite difference, extra dimension-based sparsity regularization schemes for the tasks of fluent speaking, and repetitive speaking. These improvements are attributed to the non-local soft weighted temporal regularization enabled by the proposed approach. Analysis of the mechanics of the Laplacian matrix revealed implicit motion binning enabled by the manifold regularization. Region of interest pixel time profile analysis during production of target speech sounds demonstrated improved vocal tract kinematic depictions with the manifold scheme compared to existing schemes. Through blinded image quality evaluation from three experts in voice production and modeling research, proposed manifold regularization approach outperformed existing schemes in the categories of spatial blurring and temporal blurring. The proposed scheme did not show differences against temporal finite difference scheme in the alias artifact category, but had significant differences in this category compared to remaining schemes.

This feasibility study has few noteworthy limitations. Firstly, we did not simultaneously acquire audio. Concurrent audio acquisition is typically needed to analyze timing events during speech production, and we plan to acquire concurrent audio in future studies. Secondly, our sequence had a low time duty cycle despite using short readout spirals: 4.4 ms of the 5.8 ms TR was spent in slice-selective excitation and gradient spoiling. In the future, we will explore time-optimized slice select excitation pulse design to improve the duty cycle. Thirdly, we acquired slices only in the sagittal orientation. In the future, we will explore concurrent analysis of slices in arbitrary orientations to analyze vocal tract shaping in flexible planes. We performed our analysis on limited set of subjects (3 male and 2 female) performing fluent counting and repetitive production of consonant and vowel sounds. Further analysis with regards to diverse speakers with variations in vocal tract geometry, different native language dialects are needed to fully generalize our approach.

We implemented the low rank model using a nuclear norm regularized constraint. However, in dynamic speech MRI, the low-rank model has been implemented by leveraging a hybrid 3D k-t under-sampled spiral navigator and Cartesian based readout sequence in the partial separability model [20,21]. We did not compare our approach against this scheme as our goal was to compare reconstruction constraints using a fixed acquisition scheme, and a neck-to-neck comparison of the proposed approach to the hybrid 3D k-t based scheme is beyond the scope of this work.

Coil element numbers 11,12,13 captured significant artifact energy, which originated from portions of anatomy far outside the typical upper and infra-glottic airway FOV of interest. Recently, this was shown to be due to a combination of gradient non-linearity, spiral sampling, and subject's anatomy [43]. An approach to mitigate these artifacts was also proposed. In the future we will explore this approach to effectively include measurements from all of our coil elements.

The proposed manifold regularization scheme can be improved in a number of ways. Firstly, our implementation was not optimized for reconstruction speed. Future work will include leveraging emerging Python libraries and efficient code optimization on the GPU. Secondly, we consistently observed faint alias artifacts with our scheme. This is attributed to using a $l_2$ norm regularization on the L matrix, which has a well-known limitation of its inability to fully penalize artifact energy. To address this, we will explore the use of a $l_1$ penalty on the L matrix in the future. Thirdly, we will explore the combined benefit of employing additional spatial regularization. Finally, our manifold model assumed mapping of image frames as points on a low dimension manifold living in a high dimension ambient space. In the future, we will explore mapping smaller spatial patches from the image frames onto the higher dimension ambient space. This may be suited to model certain vocal tract shaping tasks where some articulators move at a rapid pace compared to the others.

## CONCLUSION

In this study we proposed a novel scheme for improved rapid dynamic speech MRI at 3 Tesla. Our scheme leveraged parallel imaging capabilities of a custom 16-channel vocal tract coil, short readout duration and self-navigating capabilities of variable density spirals, and implicit motion binning capabilities of manifold regularization. We achieved single slice imaging at 17.2 ms/frame and concurrent 3-slice imaging at 52.2ms/frame at a spatial resolution of 2.4 mm$^2$. We demonstrated applicability of our scheme to image repetitive and fluent speaking tasks. Our approach demonstrated increased fidelity in capturing speech motion patterns compared to existing view-sharing, low rank, or sparsity-based reconstruction constraints.

**ACKNOWLEDGMENT**

This work was conducted on an MRI instrument funded by the NIH under 1S10OD025025-01.

**DATA AND CODE AVAILABILITY STATEMENT**

Example data and code from this work are available at: https://github.com/rushdi-rusho/manifold_speech.

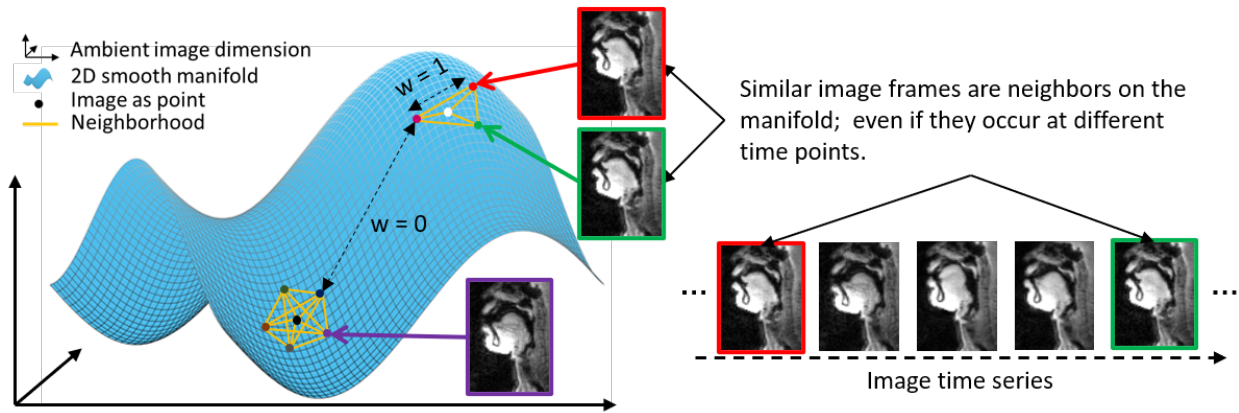

*Figure 1:* Dynamic images can be modeled as points on a smooth nonlinear manifold embedded in a high dimensional ambient space. This is demonstrated in this schematic using image frames from a fluent speaking task. Image frames sharing similar vocal tract postures are mapped as neighbors on the 2D manifold even if they occur at different times (see red and green squares). Dissimilar images are distant on the 2D manifold even if they occur consecutively in time. Weights (w) between frames determine the degree of similarity and are inversely proportional to the distance of the points on the manifold. The manifold regularization exploits the neighborhood relations of points on this manifold using a penalized optimization framework.

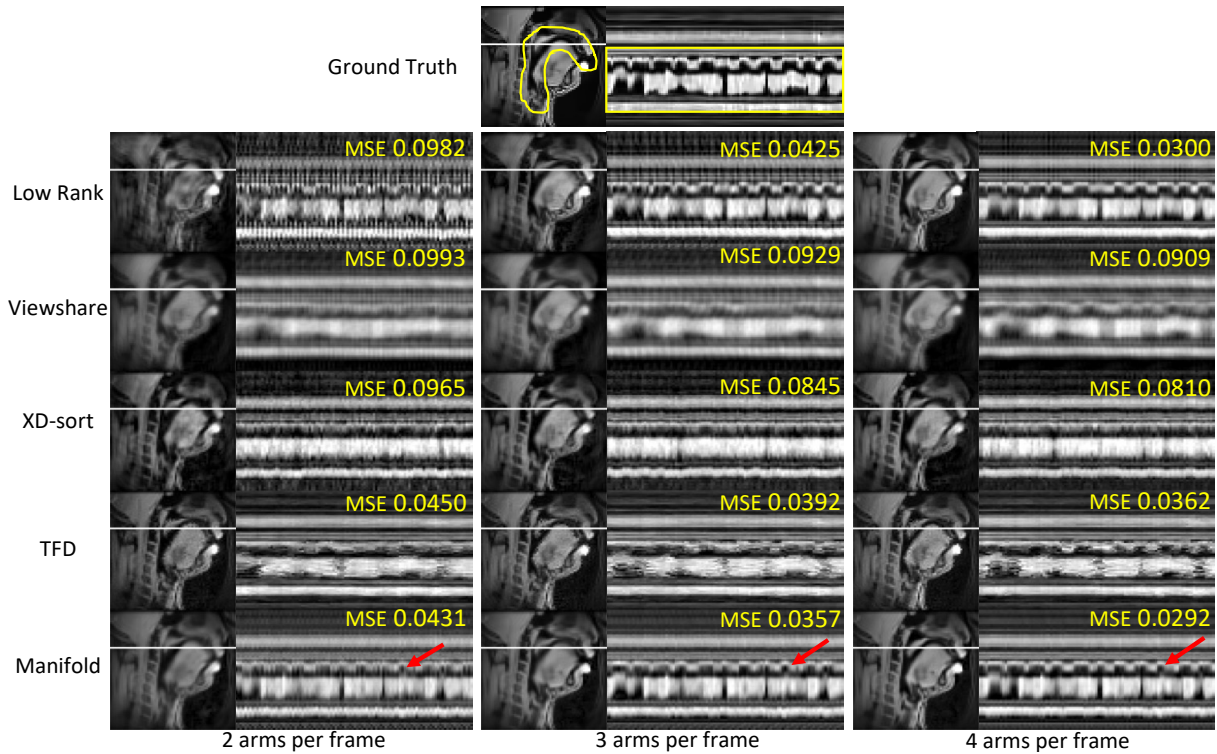

*Figure 2:* Reconstruction comparison of various algorithms on the retrospectively under-sampled data on first eighteen seconds of data where the subject was counting numbers indefinitely. The top row shows the ground truth fully-sampled data. Example spatial frame and the image time profile along the horizontal white line is shown. Comparisons are shown at different under-sampling factors corresponding to using 2,3, and 4 spiral arms/frame. The mean square error (MSE) between the reconstruction and the ground truth were evaluated in the yellow marked region of interest containing the moving air-tissue boundaries of the vocal tract. Note, the low-rank reconstruction show pronounced unresolved aliasing in the 3 and 2 arms/frame setting. The view-sharing, XD-sort algorithms demonstrate substantial temporal blurring. TFD show classic temporal stair casing motion artifacts. In contrast, the proposed manifold regularized reconstruction provides improved spatial and temporal fidelity in comparison to existing algorithms.

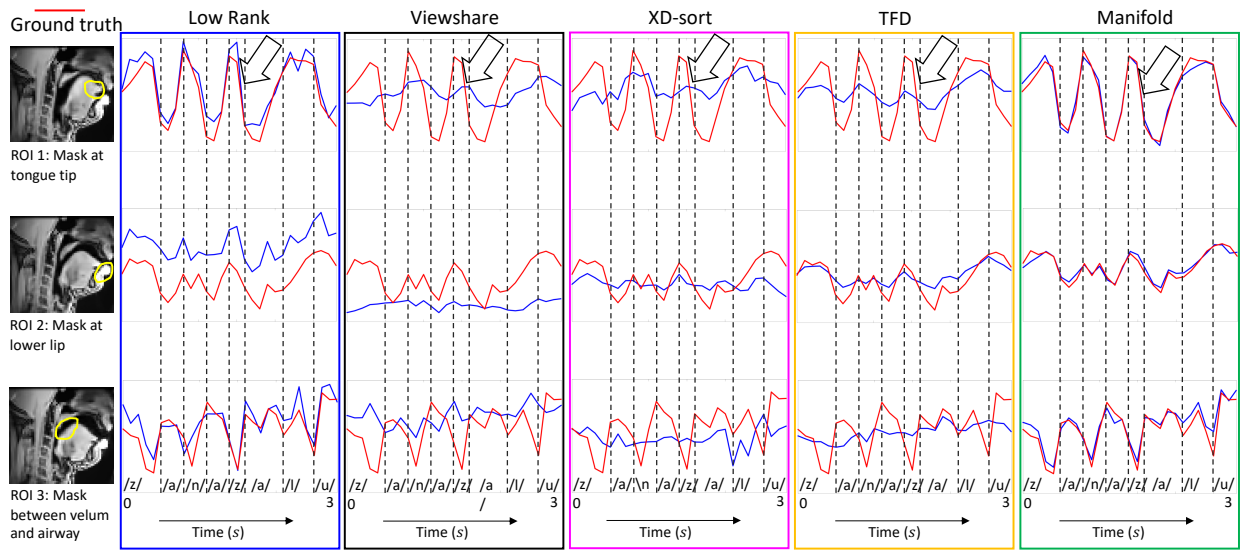

*Figure 3:* Region of interest (ROI) pixel time intensity analysis on the 3 arm/frame retrospectively under-sampled data. Three ROI masks are considered: (top row): ROI at the vocal tract air-space and tongue top; (middle row): ROI at the vocal tract airspace and lower lop; (bottom row): ROI at the vocal tract air-space and velum. The averaged intensity of pixels in every ROI is plotted as a function of time for the ground truth (red) and the various reconstruction algorithms (blue). The peaks and troughs of the profiles indirectly capture the soft tissue moving in and out of the ROIs. Note, the improved alignment of the manifold-based ROI time profiles against the ground truth in comparison to existing algorithms. For example, see the white arrows showing the tongue tip closure during the production of sound /z/.

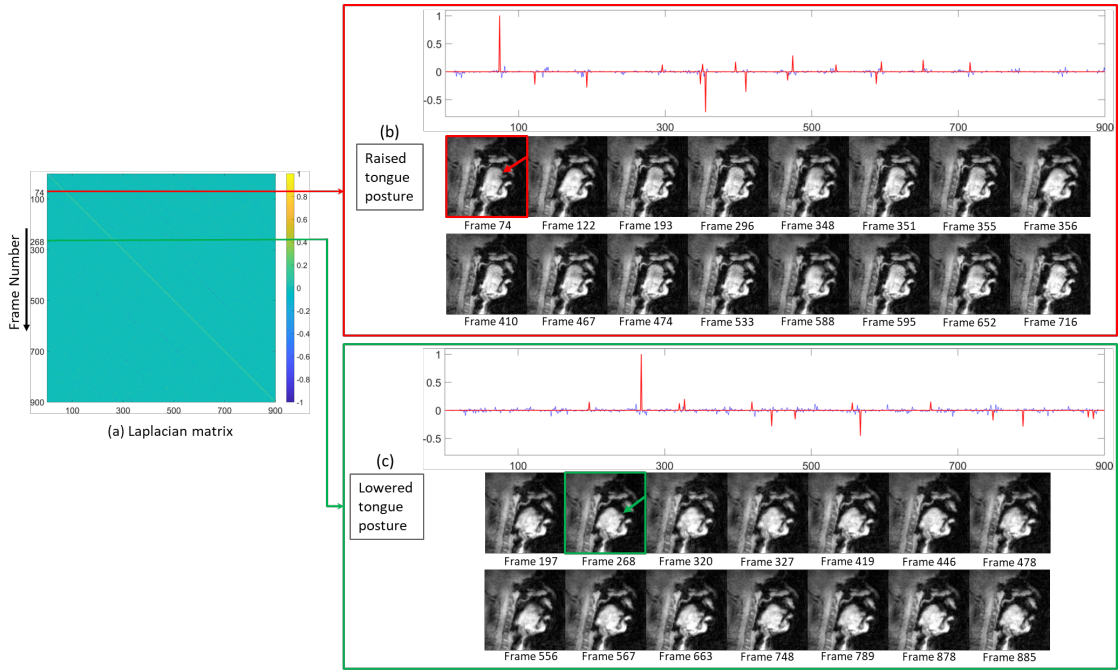

*Figure 4:* Visualizing the graph Laplacian matrix for the speaking task of repeating the phrase "loo-lee-la-za-na-za". Representative midsagittal reconstruction from speaker 1 imaged with the 3-slice acquisition scheme was considered. (a) shows the estimated Laplacian matrix. (b) shows a representative row (row # 74) of this matrix, where entries that are greater than 10% of the maximum (or) minimum value of that row are highlighted in red color, and superimposed on the remaining entries in blue color. The 74$^{th}$ frame of this sequence had a raised tongue posture as seen in the highlighted red box in (b). While reconstructing this frame, similarity between frames corresponding to peaks in the 74$^{th}$ row of the L matrix are implicitly leveraged. These frames are shown in (b), and clearly depict the tongue raised posture. Similarly, (c) shows a representative row (row # 268) of this matrix, which corresponds to the 268$^{th}$ frame depicting a lowered tongue posture (see green box in (c)). Also, note how the peaks in (c) corresponded to image frames sharing the lowered tongue posture. We also note quasi-periodic peaks in the rows representative of the repetitive nature of the speaking task.

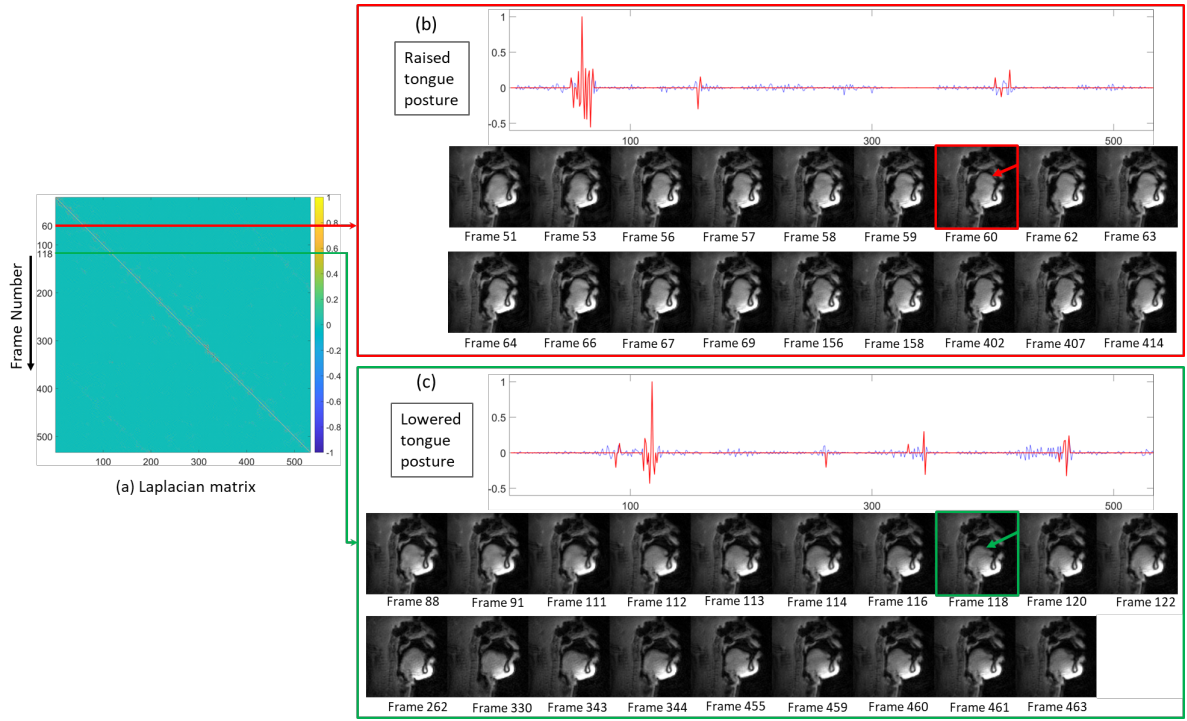

*Figure 5:* *Visualizing the graph Laplacian matrix for the speaking task of fluently counting numbers indefinitely. Representative midsagittal reconstruction from speaker 1 imaged with the single-slice acquisition scheme was considered. (a) shows the estimated Laplacian matrix. (b) shows a representative row (row # 60) of this matrix, where entries that are greater than 10% of the maximum (or) minimum value of that row are highlighted in red color, and superimposed on the remaining entries in blue color. The 64th frame of this sequence had a raised tongue posture as seen in the highlighted red box in (b). Similar to figure 4, we note similar frames that occur non-locally are identified by peaks of this row. Similarly, (c) shows a representative row (row # 118) of this matrix, which corresponds to the 118$^{th}$ frame depicting a lowered tongue posture (see green box in (c)), and note the similar frames sharing this posture being highlighted by the peaks. In contrast to Figure 4, we note the peaks in this task were arbitrary, which is representative of the fluent counting speaking task.*

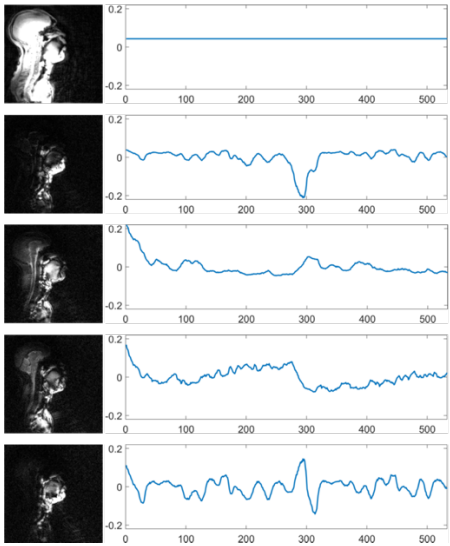
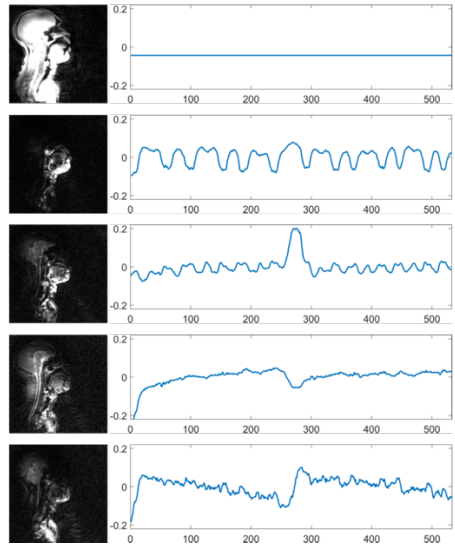

*Figure 6:* *Illustration of temporal bases (V) or the eigen vectors of the Laplacian matrix; and the spatial coefficients (U) for the speaking tasks of fluently counting numbers and repeating the phrase "za-na-za". For brevity, the first 5 bases and coefficients are shown. Note quasi periodic patterns are captured in the temporal bases of the repetitive "za-na-za" task, and more arbitrary patterns are captured in the temporal bases of the fluent speaking task.*

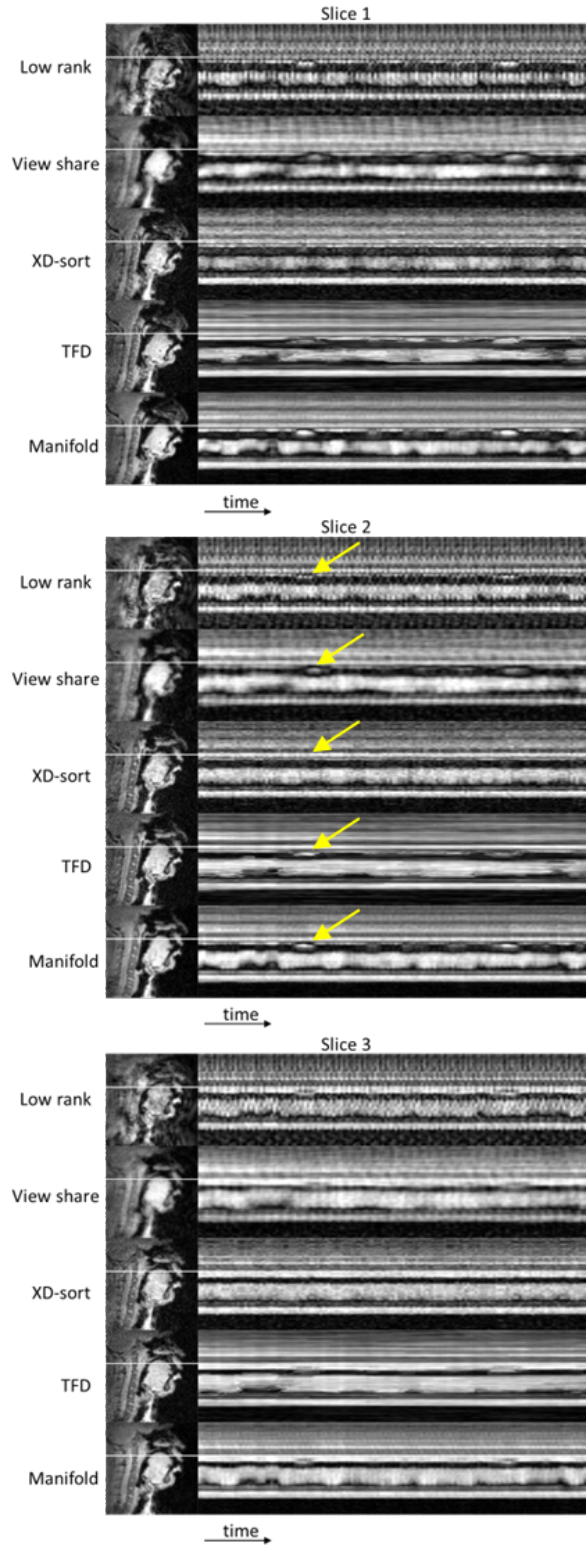

*Figure 7:* Qualitative comparison of reconstructions from low rank regularization, view-sharing, XD-sort temporal finite difference regularization, and the proposed manifold regularization schemes.

*Reconstructions were performed from the concurrent 3-slice prospectively accelerated acquisition scheme with 3arms/frame, and a native time resolution of ~52.2 ms/frame. The speaker was producing the fluent task of counting numbers indefinitely. Shown are one frame of the reconstruction along with the temporal profile cuts (indicated by the white horizontal line). Motion blurring and residual alias artifacts were present in the low-rank scheme. Temporal finite difference scheme showed no aliasing artifacts, but had considerable inter-frame motion artifacts, and temporal stair casing artifacts in the temporal profiles. XD-sort and view-sharing showed significant temporal blurring. The manifold scheme provided superior spatial and temporal fidelity compared to competing algorithms as evident by crisper temporal profiles (see arrows).*

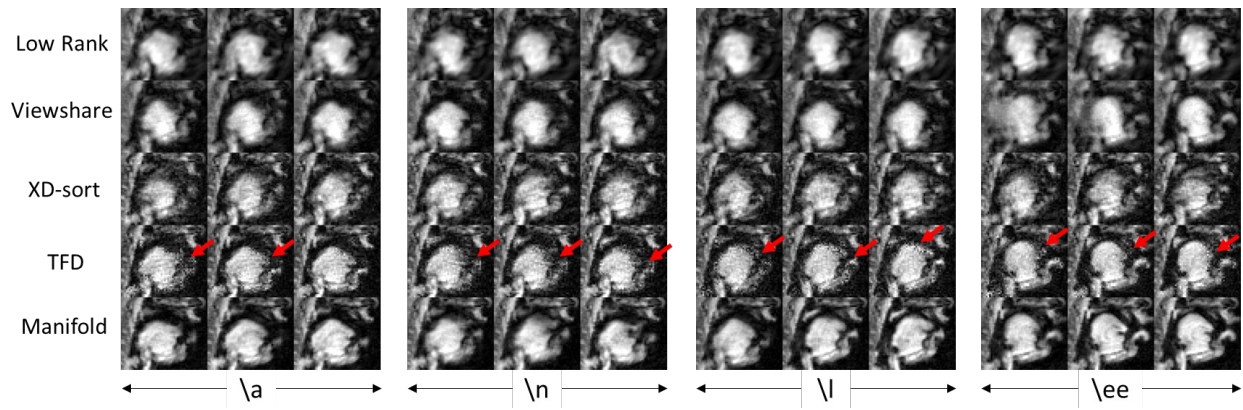

*Figure 8: Qualitative comparisons of reconstructions in the prospectively accelerated 3-slice acquisition scheme while the subject was producing repetitions of the phrase "loo-lee-la-za-na-za". We show example mid-sagittal image frames during the transition of the vocal tract in and out of the /a/, /n/,/l/, /i/ sounds. Note, the proposed manifold regularized scheme depicts clearer depiction of the vocal tract air-tissue boundaries, compared to competing schemes. For example, see the velar boundaries in /i/ sound, tongue tip boundaries in the /n/ sound, overall tongue body configuration in the /a/ sound.*

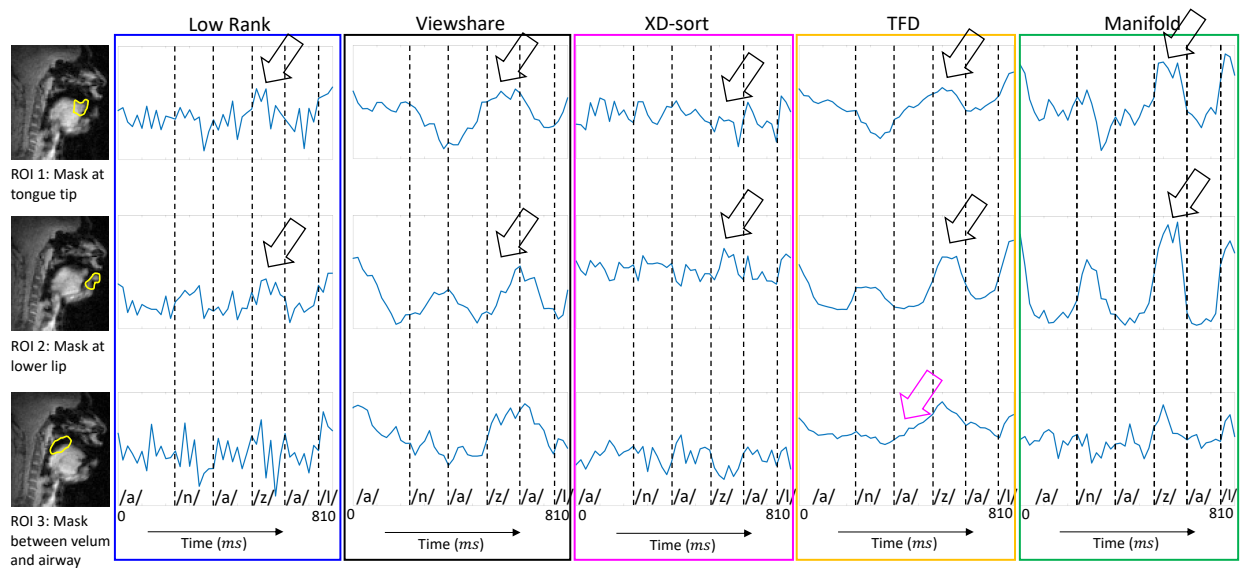

*Figure 9: Region of interest (ROI) time profile analysis on the mid-sagittal reconstructions from the concurrent 3-slice dataset for the speaking task of producing the phrase "loo-lee-za-na-za". We particularly zoom into image frames producing the sounds /a/-/n/-/a/-/z/-/a/-/l/. Three different ROIs were considered: airway near the tongue tip (ROI1), airway near the lower lip (ROI2), and airway behind the tongue and in front of the velum (ROI3). Mean pixel intensities in each of these ROIs are plotted as a function of time. The tongue tip in ROI 1 and the lower lips in ROI 2 should be in the raised position at the beginning of the sounds /n/, /z/ and /l/ which should be reflected as a sharp increase of ROI time profile. This behavior is represented well in the manifold reconstruction but not in the other schemes (eg. see black arrows). Similarly, the area between velum and airway in the ROI 3 should not change much for the sound /a/ which should be depicted as flat lines in the ROI- averaged time profiles, and is represented well in the manifold scheme compared to temporal finite difference scheme (see magenta arrow).*

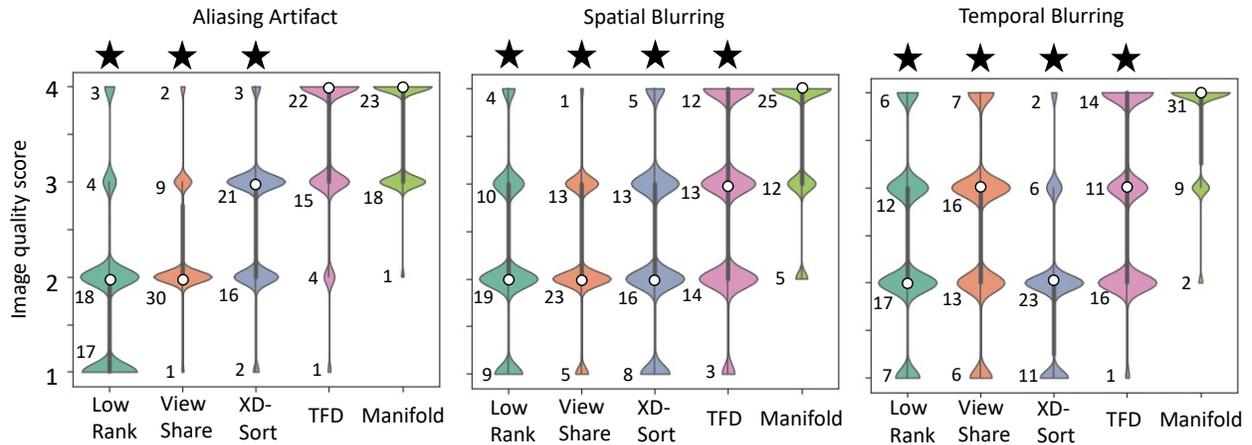

*Figure 10:* Combined image quality scores from three expert raters. The score distribution across the 14 datasets from the 3 raters are shown as violin plots for the categories of alias artifacts, spatial blurring, and temporal blurring. The median of the scores is indicated by the white circle, and the interquartile range is indicated by the black vertical box. The density of the violin plot at a particular score is proportional to the number of times that score was assigned (also listed as a number in the plot). We observed that the proposed manifold scheme consistently provided scores in the 3's (good quality) and 4's (excellent quality) all categories in comparison to the other schemes. Statistically significant differences were observed in the image quality scores with the manifold reconstruction scheme compared to all schemes in the spatial blurring, temporal blurring categories ($p<0.001$). In the aliasing artifact category, except for the differences between TFD and manifold schemes, the manifold scheme showed superior scores compared against other methods ($p<0.001$).